\documentclass[a4paper]{spie_margins}

\usepackage{amsmath,amsfonts,amssymb}
\usepackage{graphicx}
\usepackage[colorlinks=true, allcolors=blue]{hyperref}

\usepackage{textcomp}    
\usepackage[dvipsnames]{xcolor}      
\usepackage{aas_macros}  
\usepackage{multirow,makecell}

\graphicspath{{./images/}}  

\title{Developing the GOTO Telescope Control System}

\author[a]{\mbox{Martin J. Dyer}}
\author[a,b]{\mbox{Vik S. Dhillon}}
\author[a]{\mbox{Stuart Littlefair}}
\author[c]{\mbox{Danny Steeghs}}
\author[c]{\mbox{Krzysztof Ulaczyk}}
\author[c]{\mbox{Paul Chote}}
\author[c]{\mbox{Joseph Lyman}}
\author[d]{\mbox{Duncan K. Galloway}}
\author[d]{\mbox{Kendall Ackley}}
\author[d]{\mbox{Yik Lun Mong}}
\author[ ]{\mbox{the GOTO Collaboration}}

\affil[a]{Department of Physics and Astronomy, University of Sheffield, Sheffield S3 7RH, UK}
\affil[b]{Instituto de Astrof\'{i}sica de Canarias, E-38205 La Laguna, Tenerife, Spain}
\affil[c]{Department of Physics, University of Warwick, Coventry CV4 7AL, UK}
\affil[d]{School of Physics \& Astronomy, Monash University, Clayton VIC 3800, Australia}

\authorinfo{Send correspondence to MJD - email: martin.dyer@sheffield.ac.uk}

\begin{document} 
\maketitle

\begin{abstract}
The Gravitational-wave Optical Transient Observer (GOTO) is a wide-field telescope project focused on detecting optical counterparts to gravitational-wave sources. The GOTO Telescope Control System (G-TeCS) is a custom robotic control system which autonomously manages the GOTO telescope hardware and nightly operations.
Since the commissioning the GOTO prototype on La Palma in 2017, development of the control system has focused on the alert handling and scheduling systems. These allow GOTO to receive and process transient alerts and then schedule and carry out observations, all without the need for human involvement.
GOTO is ultimately anticipated to include multiple telescope arrays on independent mounts, both on La Palma and at a southern site in Australia. When complete these mounts will be linked to form a single multi-site observatory, requiring more advanced scheduling systems to best optimise survey and follow-up observations.
\end{abstract}


\section{The GOTO project}
\label{sec:goto}

\begin{figure}[t]
    \begin{center}
        \includegraphics[width=0.49\linewidth]{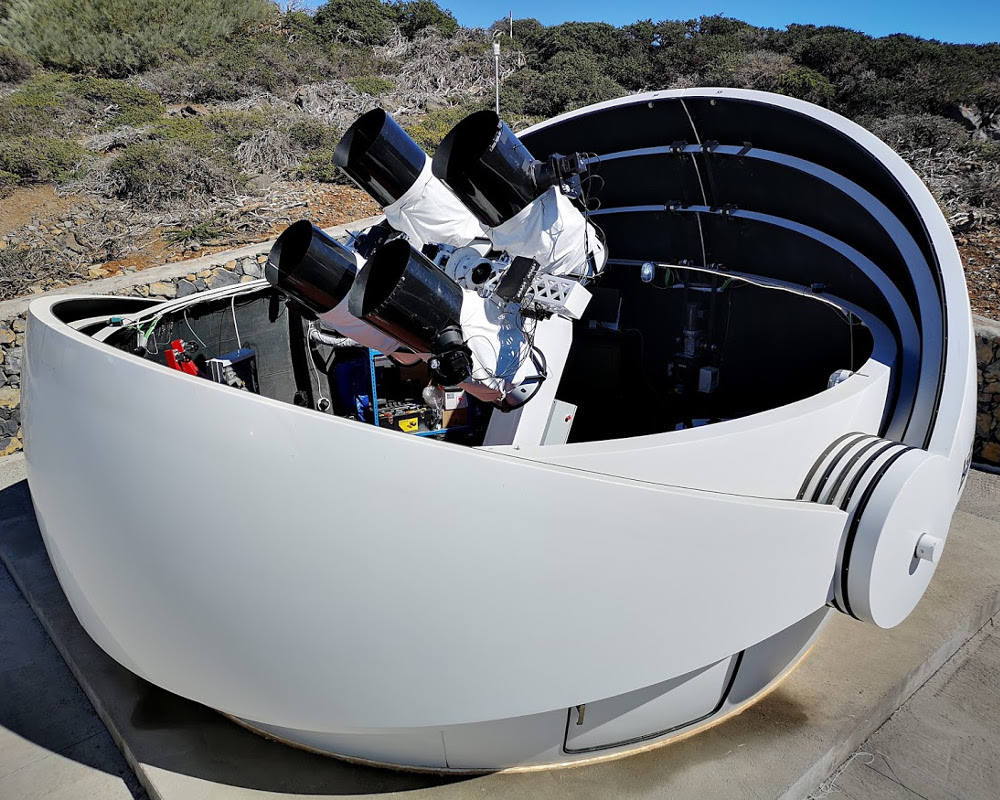}
        \includegraphics[width=0.49\linewidth]{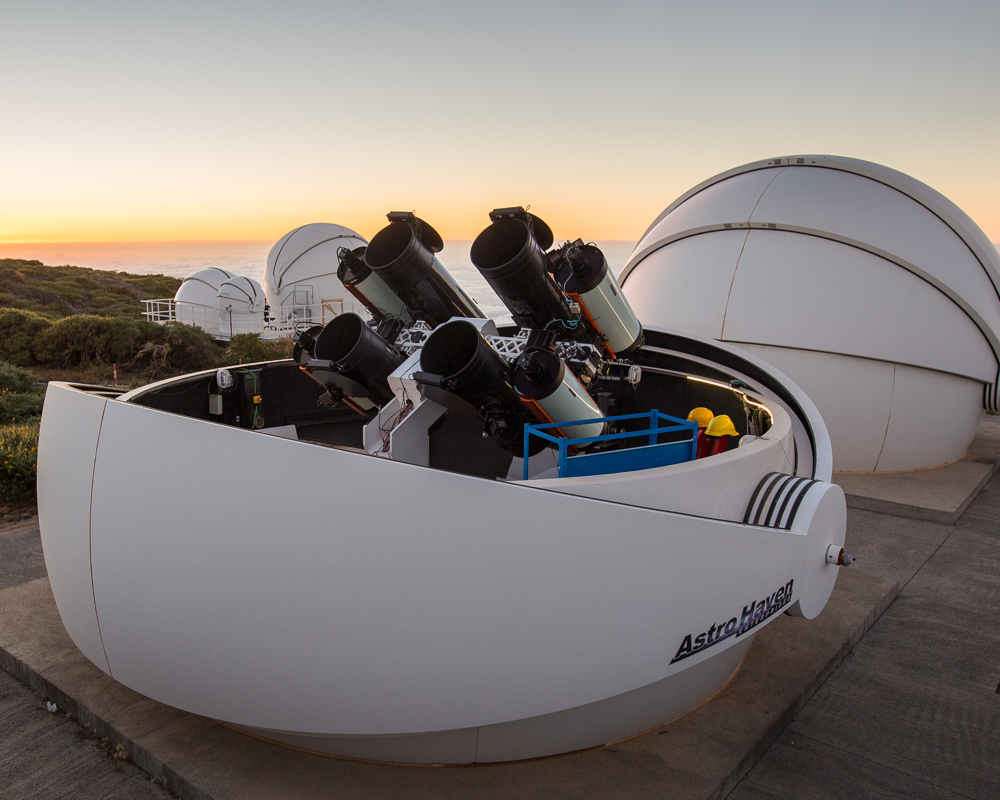}
    \end{center}
    \caption{The GOTO prototype on La Palma: with the initial four unit telescopes (left) and after the 2020 upgrade (right). The four outer tubes shown on the right are temporary stand-ins while the original OTAs are refurbished.}
    \label{fig:goto_photo}
\end{figure}

The Gravitational-wave Optical Transient Observer (GOTO) collaboration\footnote{\url{https://goto-observatory.org/}} was founded in 2014, with the aim of detecting optical counterparts to gravitational-wave sources through the construction of a network of robotic, wide-field telescopes\cite{GOTOprototype}. Each GOTO telescope consists of an array of unit telescopes (UTs) attached to a single, fast-slewing robotic mount, and with a complete array of eight UTs each GOTO telescope will have an instantaneous field of view of $\sim$40~square~degrees. Each GOTO site will ultimately host two independent mounts in separate domes. Under normal operations each telescope will carry out an all-sky survey, building up an archive of deep reference images. When a gravitational-wave alert is received the telescope will switch to covering as much of the visible probability skymap as possible, using the same grid as the all-sky survey. Through the process of difference imaging, new transient sources will be detected by comparing the new images to the reference image taken at the same location.

The first GOTO prototype telescope was constructed at the Observatorio del Roque de los Muchachos on La Palma in the Canary Islands in 2017, and featured four 40~cm unit telescopes (Figure~\ref{fig:goto_photo}, left). In 2020 the prototype unit telescopes were replaced with new, second-generation models with enclosed tubes (Figure~\ref{fig:goto_photo}, right). An additional mount with eight unit telescopes is under construction, and will be deployed to a second neighbouring dome on La Palma in early 2021. Funding has also been secured for a southern node at Siding Spring Observatory in New South Wales, Australia, which will host an additional two domes each with independent mounts and eight UTs. When complete the GOTO network will be able observe and react to alerts with near 24-hour coverage, with an instantaneous field of view of $\sim$80~square degrees at each site when utilising both mounts. It is planned that the entire network will be operational by the time that the LIGO-Virgo-KAGRA gravitational-wave detectors resume operations in 2022\cite{LIGO-Virgo-KAGRA}.

\newpage

\section{Robotic telescope control}
\label{sec:control}

A custom software package called the GOTO Telescope Control System (G\nobreakdash-TeCS)\footnote{\url{https://github.com/GOTO-OBS/g-tecs}} was developed to control all operations of the robotic telescope.\cite{gtecs,thesis} G\nobreakdash-TeCS controls all aspects of the telescope, including the physical hardware, observation scheduling and automated operations such as ensuring the dome closes in poor weather conditions. Initial development of G\nobreakdash-TeCS focused on the hardware control software and the pilot, and was described previously in [\citeonline{gtecs}]. Sections~\ref{sec:hardware}~and~\ref{sec:autonomous} summarise the previous work, before detailing the more recent developments in the alert follow-up system in Sections~\ref{sec:scheduling}~and~\ref{sec:transients}.

\subsection{Hardware control}
\label{sec:hardware}

\begin{figure}[t]
    \begin{center}
        \includegraphics[width=0.85\linewidth]{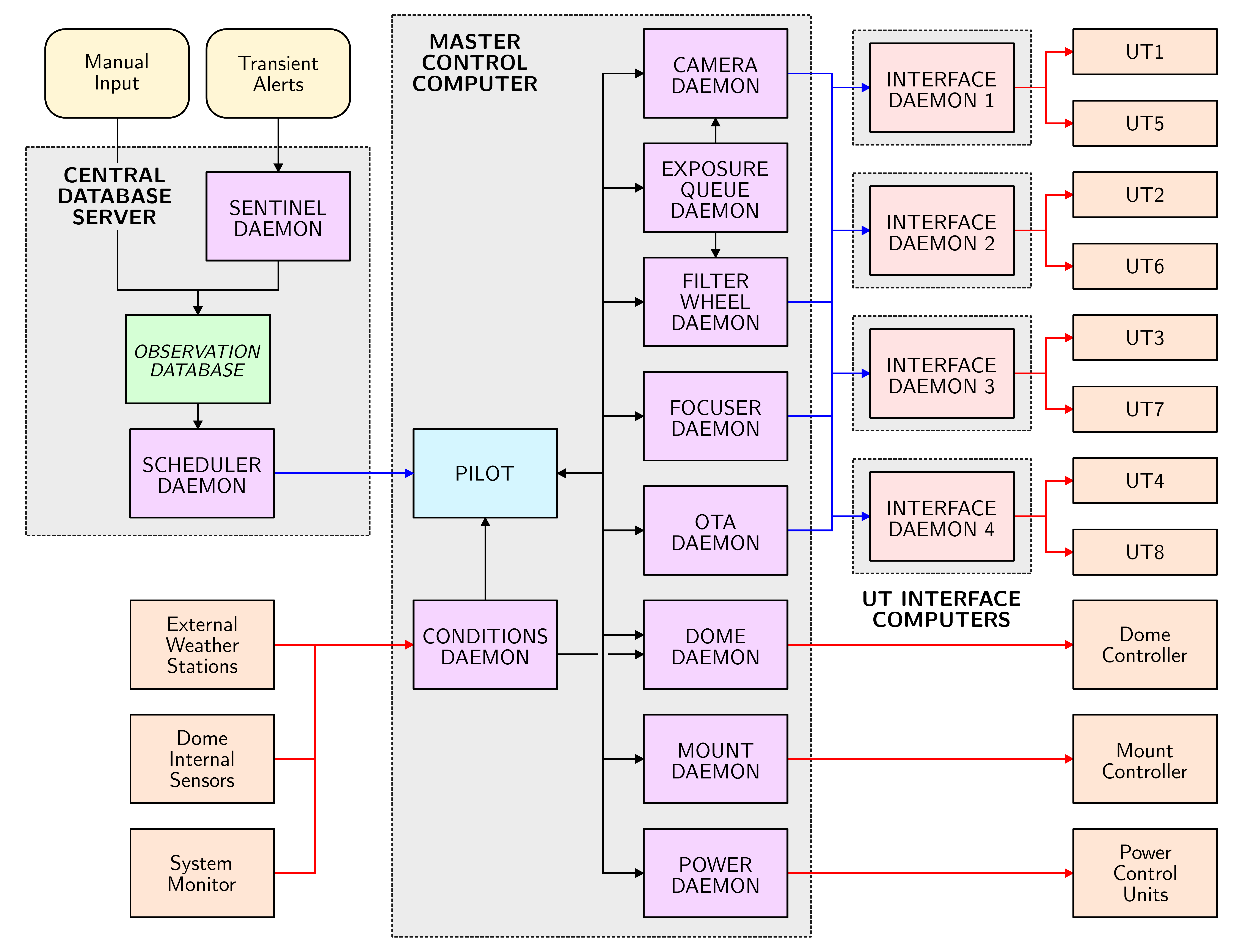}
    \end{center}
    \caption{The G-TeCS software architecture.}
    \label{fig:gtecs_flow}
\end{figure}

Each category of hardware has an associated daemon, which monitors the hardware and allows commands to be sent to it. There are eight primary hardware control daemons as shown in Figure~\ref{fig:gtecs_flow}: the camera, exposure queue, filter wheel, focuser, OTA, dome, mount and power daemons. The dome and mount daemons are solely for controlling and monitoring the clamshell dome and robotic mount respectively, while the power daemon monitors and provides a unified interface for multiple networked power supply and distribution units. Due to the design of GOTO, with multiple unit telescopes on each mount, the camera, filter wheel, focuser and OTA daemons interact with their hardware units through simple interface daemons running on small Intel PCs attached to the main boom-arm. Using these interfaces the daemons each control the four appropriate hardware units (eight in the final design) as one. For example, the filter wheel daemon can change the filter of each unit telescope individually or together in parallel. Likewise exposures on each camera are started, downloaded and saved simultaneously. The exposure queue daemon provides a common interface for taking sets of multiple exposures by issuing commands to the camera daemon in sequence, and also allows for filters to be set between exposures. The OTA daemon is a recent addition to control the mirror covers on the new enclosed tubes.

The hardware daemons offer a simple interface to the physical hardware, and they regularly fetch and store any relevant status information. These daemons however are not designed to operate autonomously, and will not issue commands to the hardware unless told to by a human user or the robotic pilot (see Section~\ref{sec:pilot}). The exception to this is the dome daemon, which has a limited amount of autonomy enabling it to automatically close the dome if the weather conditions turn bad (hence the direct connection between it and the conditions monitor shown in Figure~\ref{fig:gtecs_flow}). This provides a critical backup to ensure the telescope is protected in case the order to close from the pilot is delayed or the system is in manual mode (see Section~\ref{sec:modes}).

\newpage
\subsection{Autonomous systems}
\label{sec:autonomous}

The hardware daemons are used to interface with and monitor the telescope hardware, however aside from the dome closing in bad weather they have no autonomy of their own and need to be sent commands by either a human observer or by the \emph{pilot} master control script. The pilot is an asynchronous Python script containing multiple coroutines, which together monitor and operate the telescope hardware.

\subsubsection{Pilot routines}
\label{sec:pilot}

The primary routine within the pilot is called the night marshal, and its purpose is to run predefined observing scripts at set times during the night based on the altitude of the Sun. At the start of the night the pilot powers on the hardware and takes bias and dark frames. Once the Sun has set below the horizon, and the conditions are safe to do so, the night marshal issues the command to open the dome and starts the script to take sky flats in each of the four \textit{LRGB} filters. When taking flats the telescope is pointed at the anti-sun position\cite{flats}, and exposure times are calculated based on the sky brightness\cite{flats2}. After the flats script has finished the night marshal starts the autofocus routine, which focuses each unit telescope in parallel using the half-flux diameter V-curve method\cite{autofocus}. Once the telescopes are focused the pilot switches to observing mode. For the rest of the night the pilot fetches the highest priority pointings from the scheduler (see Section~\ref{sec:scheduling}), slewing the mount to each target and taking the requested exposures, until the Sun rises the next morning. Then the night marshal runs the flats routine again to take additional twilight flats, before closing the dome and shutting down the system. For security there is an independent night countdown timer within the pilot, which will trigger the shutdown when the Sun rises above the horizon if the night marshal hasn't already done so. There is also an entirely separate script, called the day marshal, which runs every morning and ensures that the dome is closed in case of an error in the pilot.

\newpage

Aside from the night marshal, the pilot also contains multiple check routines that monitor different aspects of the control system. One routine fetches the latest pointings from the scheduler (see Section~\ref{sec:scheduling}) every 10 seconds, interrupting the current observation if the new target has a higher priority. Another routine monitors the hardware daemons for any errors, which are defined by comparing the expected status to the actual one. If the two differ (e.g. the cameras are not at the requested temperature, or dome is open when it should be closed) the pilot is able to issue a series of predefined recovery commands to the daemons to attempt to fix the problem. If this is unsuccessful then the pilot shuts down and sends out alerts for human assistance. A third routine is in charge of pausing and resuming the night marshal if required: if a hardware error is detected, the system is in manual mode (see Section~\ref{sec:modes}), or the conditions flags are bad (see Section~\ref{sec:conditions}).

\subsubsection{Observing modes}
\label{sec:modes}

Although GOTO is capable of operating fully autonomously, it is sometimes necessary for a human operator to take control. An example of the need for a human observer was when GOTO was used to observe the asteroid Phaethon\cite{Phaethon}, as the existing scheduling system was not designed for moving objects. It is also important for safety that the automated systems can be reliably disabled if there are workers or observers on site, as it could be dangerous if the system still tried to move the mount or dome unexpectedly. G\nobreakdash-TeCS therefore includes three operating modes: robotic, manual or engineering. Robotic mode is the default, and while the system is in this mode the pilot will be in complete control of the telescope and the hardware will moved autonomously (for example the dome will always automatically close in bad weather). Manual mode is intended for cases when a human is in charge of the telescope, either on-site or remotely. In this mode the pilot will be paused so it will not interrupt commands sent by the observer, and while the dome will still close automatically if the conditions are bad this, and all the automated systems, can be disabled if the observer chooses. Finally engineering mode is used when there are workers on site. In this mode all autonomous systems are disabled and cannot be enabled, and the pilot will refuse to start.

\subsubsection{Conditions monitoring}
\label{sec:conditions}

Before GOTO could be trusted to run without human supervision it was necessary for it to have a robust and reliable system for closing the dome in bad weather. G\nobreakdash-TeCS therefore contains a conditions daemon, shown in Figure~\ref{fig:gtecs_flow}, which is tasked with monitoring the site conditions.

The conditions daemon monitors standard weather values reported by the on-site weather stations located just a few metres from the GOTO domes. This includes the rain, temperature, humidity, dew point, wind speed and wind gust values. Other sources of conditions information includes the internal temperature and humidity sensors, the network link, the amount of free disk space, the power level of the dome UPS and the status of the dome access hatch. The daemon processes these inputs into a series of output flags, which have a value of \texttt{0} (good), \texttt{1} (bad) or \texttt{2} (error). Each conditions flag has a limit below or above which the flag will turn from good to bad, as well as delay times that have to be exceeded before the flag changes (therefore smoothing out any sudden spikes). If any of the flags are marked as not good (i.e.\ the sum of all flags is $>0$) then the overall conditions are bad, which will trigger the dome to close and the pilot to pause its current task.

A recent addition was a manual override flag, which can be set while the system is still in robotic mode for cases that the conditions monitor is not able to detect. This is most commonly used to pause observing during cloudy skies when the weather is otherwise good, as there is no automatic cloud sensor connected to the conditions monitor. The flag can also be used for rarer weather conditions on La Palma, such as smoke from forest fires or high levels of dust from the Sahara.

\newpage
\subsection{Observation scheduling}
\label{sec:scheduling}

\begin{table}[t]
    \begin{center}
        \begin{tabular}{cc|cccc} 
            &
            & \multicolumn{4}{c}{Highest priority pointing is\ldots}
            \\[0.1cm]
            &
            & \makecell{same as \\ current pointing}
            & \makecell{a new, \\ valid pointing}
            & \makecell{a new, \\ invalid pointing}
            & (None)
            \\[0.3cm]
            \hline
            \multirow{4}{*}{\rotatebox[origin=c]{90}{Current pointing is\ldots~~}}
            & & & & &
            \\[-0.2cm]
            & valid
            & \makecell{\textcolor{ForestGreen}{Continue} \\ \textcolor{ForestGreen}{current pointing}}
            & \makecell{\textcolor{BlueGreen}{Interrupt and start new pointing} \\ \textcolor{BlueGreen}{if it is a ToO and the current pointing is not,} \\ \textcolor{BlueGreen}{otherwise continue current pointing}}
            & \textcolor{Gray}{N/A}
            & \textcolor{Gray}{N/A}
            \\[0.7cm]
            & invalid
            & \textcolor{Red}{Park}
            & \textcolor{NavyBlue}{Interrupt and start new pointing}
            & \textcolor{Red}{Park}
            & \textcolor{Red}{Park}
            \\[0.7cm]
            & \makecell{(None)}
            & \textcolor{Red}{Park}
            & \textcolor{NavyBlue}{Start new pointing}
            & \textcolor{Red}{Park}
            & \textcolor{Red}{Park}
            \\[0.5cm]
        \end{tabular}
    \end{center}
    \caption[Actions to take based on scheduler results]{
        Actions the pilot will take based on the scheduler results. The scheduler will return the current pointing and the highest priority pointing, which may be the same or different (and either could be None, indicating the pilot is not currently observing or the queue is empty, respectively).
    }\label{tab:sched}
\end{table}

In order for the pilot to function during the night it needs to know what targets to observe. GOTO operates under a ``just-in-time'' scheduling model: rather than creating a plan at the beginning of the night of what to observe, each time the pilot queries the scheduler the priority of every valid pointing in the queue is calculated and the highest priority pointing at that time is returned. This system is very reactive to any incoming alerts, as the new pointings will immediately be included in the queue at the next check. This method also naturally works around any delay in observations due to poor conditions, unlike a fixed night plan. During the third LIGO-Virgo observing run GOTO was able to process gravitational-wave alerts and begin observations within 30 seconds of them being received\cite{Gompertz2020}.

The scheduler carries out several steps to determine the highest priority pointing. First the queue of pending pointings is fetched from the observation database, and any that are invalid (e.g. below the horizon, or too close to the Moon) are filtered out. The pointings are then sorted by several parameters. The first is the effective rank ($R$), which is the combination of the starting rank assigned when the pointing was entered into the database ($R_s$) and the number of times that pointing has been observed since ($n_\text{obs}$) multiplied by 10 ($R = R_s + 10 \times n_\text{obs}$). This parameter was chosen to account for observing large numbers of pointings for different events simultaneously, and is discussed in Section~\ref{sec:strategy}. For pointings of equal effective rank, any that are designated as targets of opportunity will be prioritised (and are able to interrupt observations that are in progress). In cases where there are still multiple pointings with equal priority then a tiebreak parameter ($V$) is calculated by combining any weighting ($W$) assigned to the pointings (typically from the enclosed skymap probability, see Section~\ref{sec:tiling}) and the airmass of the target ($X(t)$) at the time of the calculation. These are combined in the ratio 10:1 using
\begin{equation}
    V(t) = \frac{10}{11}W + \frac{1}{11}(2-X(t)).
\end{equation}
The combination of tile weighting and airmass is adjusted to maximise the chance of detecting a counterpart while not neglecting data quality. The 10:1 ratio was determined using preliminary strategy simulations detailed in [\citeonline{thesis}], and work is ongoing to further improve and optimise this algorithm.

The check routine within the pilot queries the scheduler every 10 seconds, and the scheduler returns both the highest priority pointing at the query time and the current pointing with its priority recalculated. The pilot compares the two and decides between three options: carry on with the current observation, switch to a new observation, or park the telescope (in the case that there are no valid targets). The different possible cases are summarised in Table~\ref{tab:sched}. Most of the time if the pilot is currently observing a pointing it will remain the highest priority in subsequent checks, so the pilot will continue observing it. Even if the scheduler finds that a different pointing now has a higher priority the pilot will not change targets unless the new pointing is marked as a target of opportunity and the current pointing is not. Otherwise the pilot will wait until it has finished the current pointing, mark it as complete in the database and ask the scheduler for the next target to observe.

\newpage
\subsection{Transient alert processing}
\label{sec:transients}

\begin{table}[t]
    \begin{center}
        \begin{tabular}{clll}
            GCN packet type & Source              & Notice type                  & GOTO-alert Event subclass           \\
            \hline
            61   & NASA/\textit{Swift} & \texttt{SWIFT\_BAT\_GRB\_POS}  & \texttt{GRBEvent}          \\
            115  & NASA/\textit{Fermi} & \texttt{FERMI\_GBM\_FIN\_POS}  & \texttt{GRBEvent}          \\
            150  & LIGO-Virgo          & \texttt{LVC\_PRELIMINARY}      & \texttt{GWEvent}           \\
            151  & LIGO-Virgo          & \texttt{LVC\_INITIAL}          & \texttt{GWEvent}           \\
            152  & LIGO-Virgo          & \texttt{LVC\_UPDATE}           & \texttt{GWEvent}           \\
            164  & LIGO-Virgo          & \texttt{LVC\_RETRACTION}       & \texttt{GWRetractionEvent} \\
        \end{tabular}
    \end{center}
    \caption[GCN notices recognised by the GOTO-alert event handler]{
        GCN notices and corresponding classes recognised by the GOTO-alert event handler.
    }\label{tab:events}
\end{table}

In order for targets to be observed by the pilot they must have entries defined in the observation database. These can be added manually, but for automated follow-up observations they have to be inserted whenever an alert is received. As shown in Figure~\ref{fig:gtecs_flow} this is the job of the G-TeCS sentinel daemon. The sentinel alert listener continuously monitors the transient alert stream output by the 4 Pi Sky event broker \cite{4pisky}, primarily the NASA GCN/TAN network\cite{GCN}, using functions from the PyGCN Python package\footnote{\url{https://pypi.org/project/pygcn}}. Alert information is received in the form of VOEvents \cite{voevent}, and are parsed using the voevent-parse Python package\cite{voevent-parse}. Each event is then processed using a Python package developed for the GOTO project, GOTO-alert\footnote{\url{https://github.com/GOTO-OBS/goto-alert}}.

\subsubsection{Event classes}
\label{sec:events}

GOTO-alert is designed around a single \texttt{Event} object class, and ``interesting'' events are processed using individual subclasses. There are currently three subclasses defined within GOTO-alert, as given in Table~\ref{tab:events}: gamma-ray burst (GRB) detections from \textit{Fermi} GBM\cite{Fermi_GBM} or \textit{Swift} BAT\cite{Swift_BAT} use the \texttt{GRBEvent} class, while the different stages of gravitational-wave (GW) detections from the LIGO-Virgo Collaboration \cite{LIGO-Virgo} use the \texttt{GWEvent} class (aside from retraction notices, which use a unique \texttt{GWRetractionEvent} class). Notices with a packet type not listed in Table~\ref{tab:events} are ignored, although the flexibility of the GOTO-alert code makes it easy to define new event subclasses to process them if desired. Likewise events that do have the correct packet type but are not marked as observations (e.g. test notices) are also rejected.

\subsubsection{Observing strategy}
\label{sec:strategy}

\begin{figure}[t]
    \begin{center}
        \includegraphics[width=0.9\linewidth]{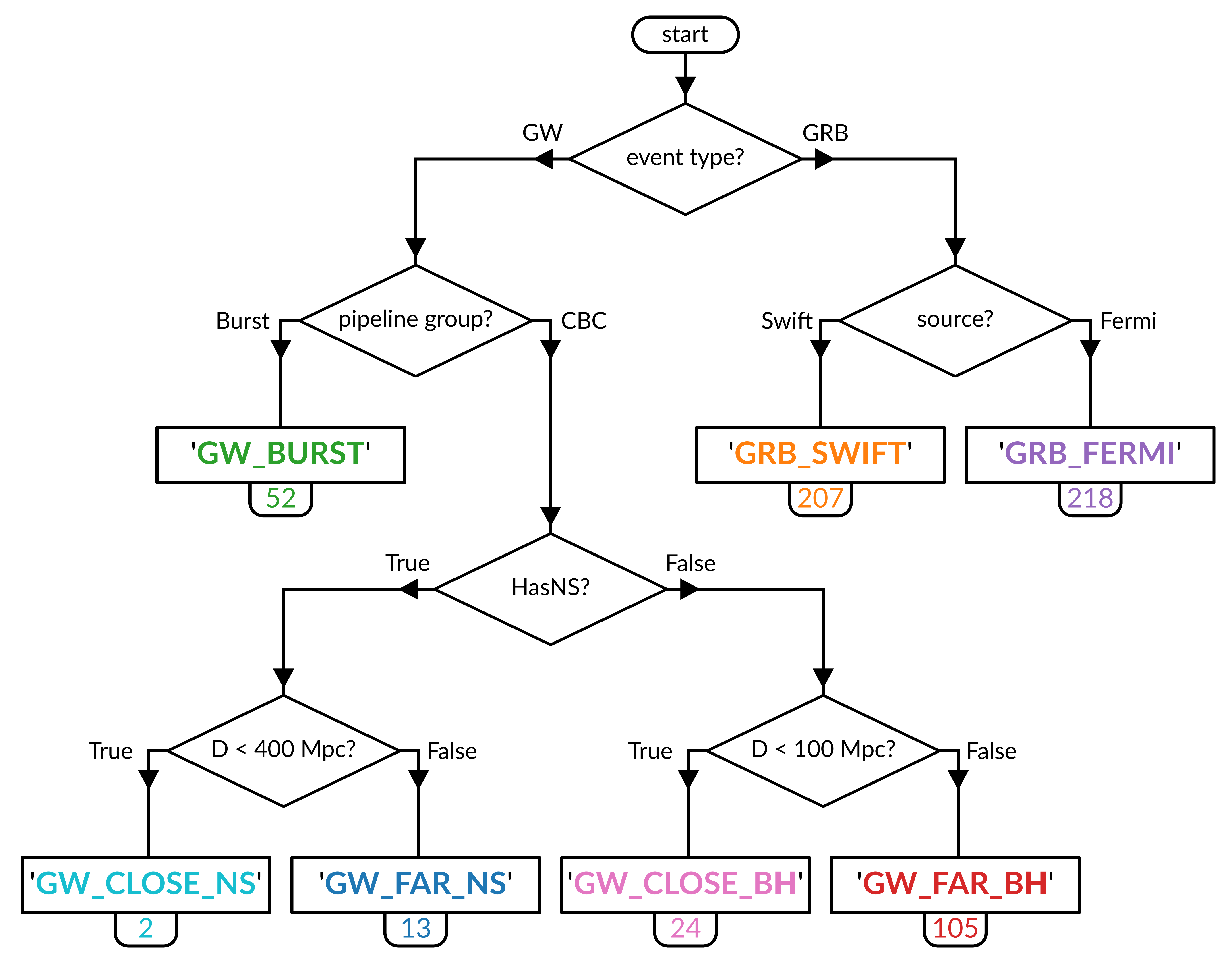}
    \end{center}
    \caption{Decision tree for determining event strategy. The number below each strategy code is the starting rank used when adding pointings for that event to the observation database (lower ranks = higher priority).}
    \label{fig:strategy_flow}
\end{figure}

Once an event has been assigned the correct subclass the sentinel passes it to the GOTO-alert event handler, which determines the correct observing strategy for the event and inserts pointings into the observation database. Observations can be tailored to the properties of the triggering event by modifying the pointing rank or observing constraints, as well as the specific exposures requested (e.g. exposure time, filters) and the desired cadence for any repeat observations. For LIGO-Virgo detections subsequent notices can contain updated localisation skymaps, overwriting those from previous alerts. In the future the observing strategy will also need to consider the different GOTO observing sites and how best to utilise the telescopes at each site (see Section~\ref{sec:future}).

To determine the priority for a given event the event handler uses the decision tree shown in Figure~\ref{fig:strategy_flow}. The codes at the end of the branches (\texttt{GW\_CLOSE\_NS}, \texttt{GRB\_FERMI}, etc) correspond to individual strategies, and event sources and parameters contained within the alert are used to decide which strategy is appropriate for a given event. First, as they are GOTO's primary science goal, gravitational-wave alerts are always prioritised over GRB alerts. Gravitational-wave signals that are detected by the CBC (Compact Binary Coalescence) pipeline can be split between those that are predicted to contain neutron stars (either neutron star-neutron star binaries or neutron star-black hole binaries) and those that are not (black hole-black hole binaries). As the former are considered more likely to produce electromagnetic counterparts they are prioritised for follow-up. Events are then further split between ``near'' and ``far'' for each branch, which is roughly determined based on the distance provided in the notice skymap. The highest priority event is therefore a gravitational-wave signal from a nearby neutron star-neutron star or neutron star-black hole binary, which is most likely to produce a counterpart that GOTO can observe. Gravitational-wave burst detections are rarer and have less certain origins, so are prioritised at a medium rank, while GRB alerts from \textit{Swift} are prioritised over those from \textit{Fermi} as they usually have smaller localisation regions.

The primary value used when scheduling pointings, as described in Section~\ref{sec:scheduling}, is the starting rank. This is assigned by the event handler and is the same for all pointing from a given event. Due to the method of sorting pointings by their effective rank, pointings from different events can differentiated and prioritised for follow-up should GOTO need to follow-up multiple events at the same time. For instance, if two alerts are received simultaneously, one from a close neutron star binary and one from a distant one, then the pointings for the first event will be inserted into the database at rank 2 while the others will be inserted with starting rank 13 (the ranks are given in Figure~\ref{fig:strategy_flow}). When the queue is sorted by the scheduler the visible pointings from the first event will always be sorted higher than the second. However, as pointings from the event are observed and are marked completed by the pilot their effective rank starts to increase: from 2 when $n_\text{obs}=0$ to 12, 22, 32 etc. Once a pointing has been observed twice its effective rank is 22, which is now below the starting rank of pointings for the second event (13). Therefore once all visible pointings from the first event have been observed twice, pointings from the second event will be sorted higher, and so will start to be observed. Subsequent pointings from this event will then also increase in effective rank from 13 to 23, 33 etc, and from then on priority will alternate between pointings from each event. It is worth emphasising that this only matters when multiple events have pending pointings in the scheduler queue at the same time, and as pointings only remain in the queue for a maximum of three days this is currently fairly rare. However, as new gravitational-wave detectors come online, and existing ones are upgraded, the requirement to decide which event to prioritise may be more common.

\newpage

\subsubsection{Mapping onto the all-sky grid}
\label{sec:tiling}

\begin{figure}[t]
    \begin{center}
        \includegraphics[width=0.9\linewidth]{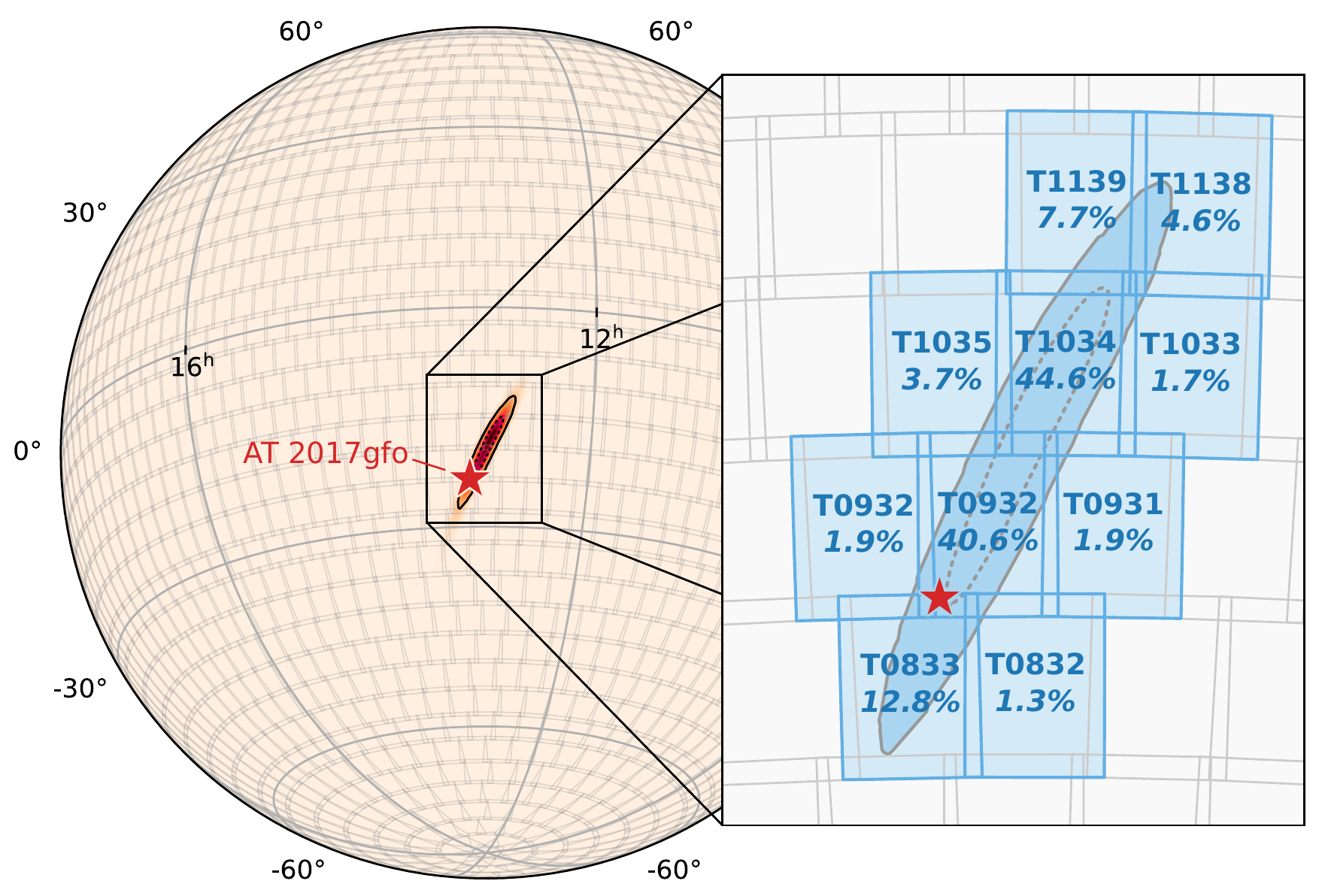}
    \end{center}
    \caption{GOTO-4 tiling applied to the skymap for GW170817. The blue tiles are those containing more than 1\% of the skymap probability. Note that overlaps between tiles means the total probability can add up to more than 100\%.}
    \label{fig:170817}
\end{figure}

GOTO carries out observations aligned to an all-sky grid, made up of tiles equal to the combined field of view of all the unit telescopes on the mount. For the GOTO-4 prototype each tile covered 3.7~degrees in right ascension and 4.9~degrees in declination, and the all-sky grid was defined by dividing the sky into a series of 2913 equally-spaced 18.1~square~degree tiles. These tiles are used as the basis of the all-sky survey that GOTO carries out when not following-up transient alerts. Survey images taken of each tile are stacked to make master reference frames, and when follow-up observations are taken aligned to the same grid new images are compared to the reference frames to detect transient sources though difference imaging (subtracting the reference from the image).

It is therefore necessary that any follow-up observations are aligned to the same grid used by the all-sky survey. LIGO-Virgo gravitational-wave notices contain probability skymaps encoded using the HEALPix framework\cite{HEALPix}, and the GOTO-alert event handler maps the skymap onto the all-sky grid by summing the values of all HEALPix pixels that are contained within each tile. An example of this mapping is shown in Figure~\ref{fig:170817}, using the initial three-detector skymap for GW170817\cite{GW170817, GW170817_followup}. Once the probability value is known for each tile, those with the highest probability values are added as pointings to the observation database, and the enclosed probability is used as the weighting factor when choosing between tiles of equal rank (see Section~\ref{sec:scheduling}). The number of tiles to add as pointings needs to be limited to ensure that tiles are revisited regularly: with the ``effective-rank'' method outlined in Section~\ref{sec:strategy}, the second observation of a tile, regardless of the probability contained within, will only take place once all other valid pointings added at the same rank have also been observed at least once. The GOTO-alert event handler selects tiles with a mean contour value of 90\% (i.e. the mean contour value of all the pixels contained within is 90\% or more). However, more quantitative simulations of different skymaps could be used to determine if 90\% is the optimal choice for all cases. For example, the contour value limit could be modified depending on the overall size of the skymap, and as more GOTO telescopes are built the limit could be lowered to add more tiles to the database, as more telescopes will be able to cover the skymap faster.

\newpage
\section{Future expansion}
\label{sec:future}

\begin{figure}[t]
    \begin{center}
        \includegraphics[width=0.95\linewidth]{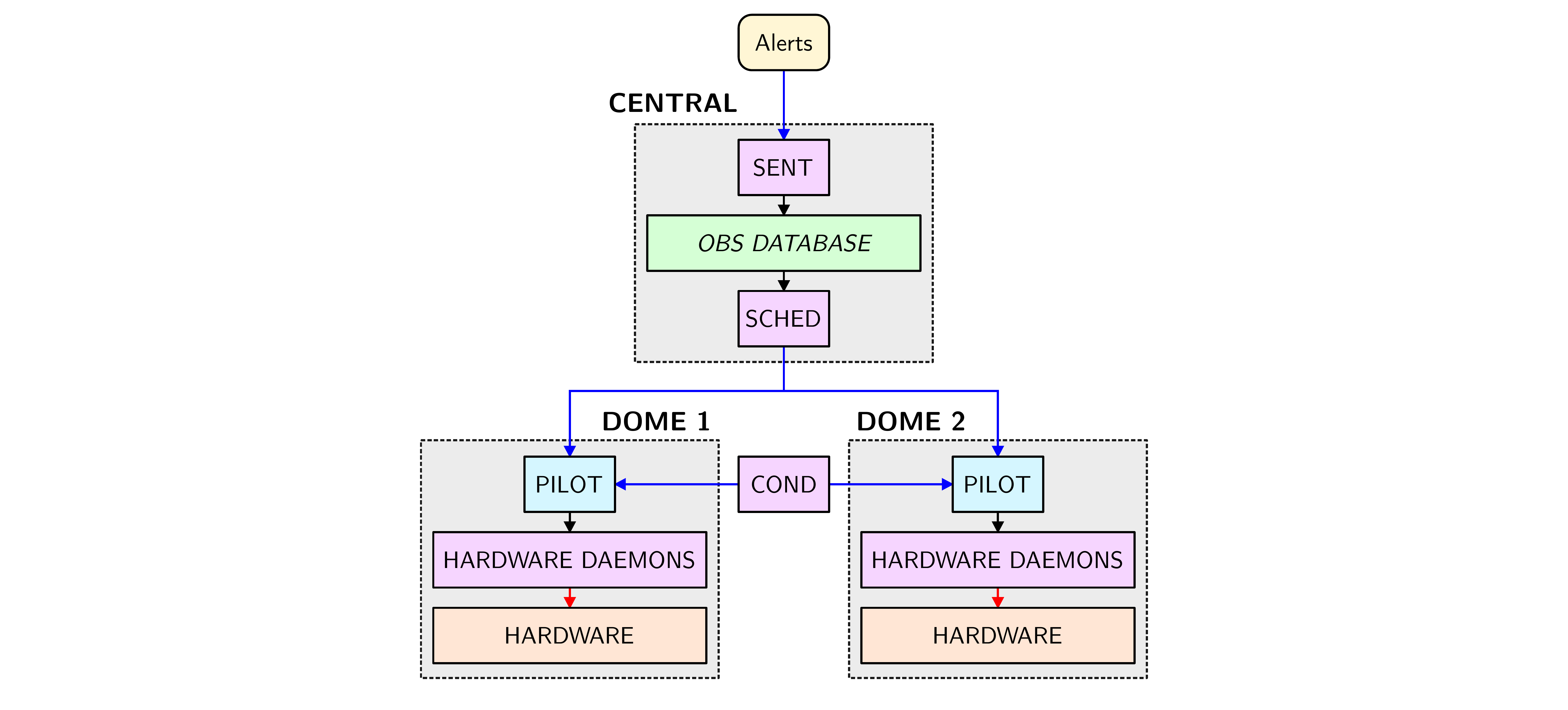}
    \end{center}
    \caption{The proposed future architecture for G\nobreakdash-TeCS, expanding Figure~\ref{fig:gtecs_flow} to cover multiple telescopes.}
    \label{fig:gtecs_flow_future}
\end{figure}

The first GOTO prototype telescope has been successfully observing on La Palma for several years, using the software systems described in Section~\ref{sec:control} to automate every aspect of its operations. During 2021, several new GOTO telescopes will be constructed: first a second mount on La Palma next to the existing telescope, then two more mounts at Siding Spring Observatory in Australia. As part of the commissioning work, G\nobreakdash-TeCS will be expanded to control the new hardware, ultimately operating all four mounts as a single, global, autonomous observatory.

\subsection{Controlling multiple telescopes}
\label{sec:multitel}

The addition of the second telescope on La Palma will require the expansion of the existing G\nobreakdash-TeCS software. The general architecture, as shown in Figure~\ref{fig:gtecs_flow}, with the pilot receiving targets from the scheduler and sending commands to the hardware daemons, will remain the same for both telescopes. Aside from a new direct-drive mount, the hardware will be the same as the current system on La Palma, so the hardware control daemons will be copied across with minimal changes. As shown in Figure~\ref{fig:gtecs_flow_future}, it is planned that each telescope will have an independent pilot, and will therefore carry out normal operations completely independently. This will ensure that, for example, a hardware fault with one telescope will not prevent the other from observing. It is also anticipated that the two pilots will share a single conditions monitor as, aside from the internal sensors in each dome, both telescopes will be monitoring the same weather stations and other system properties described in Section~\ref{sec:conditions}.

The biggest opportunity presented by having multiple connected telescopes, and therefore the most important change to the overall control system, is having them share a common scheduling system. This was always planned to take the form of each pilot querying a central scheduler, which is one of the reasons the existing system already has the sentinel, database and scheduler running on a separate server to the rest of the control system (as shown in Figure~\ref{fig:gtecs_flow}). Figure~\ref{fig:gtecs_flow_future} shows the natural extension of this, with both pilots receiving targets from the same scheduler daemon. This architecture allows the pilots to continue to operate independently, without any knowledge of the other, while any decisions relating to operating the telescopes in parallel are taken at the scheduler level.

There are numerous opportunities and challenges in developing this multi-telescope scheduling system. As described in Section~\ref{sec:scheduling}, every few seconds the existing scheduler takes the current queue and recalculates the highest priority pointing, which is then sent to the pilot. Initially, this system could be expanded relatively easy by having the scheduler simply calculate the top two highest priority pointings, and passing them on to the two telescopes so that one always observes the highest-priority pointing and the other always observes the second-highest. However in reality truly independent telescopes will never remain perfectly in sync, as the slew time taken to move to each target will not be the same, and so the scheduler will have to be able to handle asynchronous requests from either telescope at any time. The upgraded scheduler could also take the time to slew from the current position, as well as any time remaining in the current observation, into account when assigning targets. In addition, while most scheduling constraints will be the same for both mounts (e.g.\ the Moon phase or Sun altitude) it is possible that the two telescopes could have different artificial horizons and therefore altitude limits, thereby forcing the choice if the highest priority target is only visible by one telescope.

Having multiple mounts also opens up opportunities for more advanced observing strategies, which would have to be determined by the scheduler. The two telescopes could both observe different parts of the sky to cover the survey or gravitational-wave skymaps faster, or they could observe the same tiles to achieve a greater depth when the images are stacked. The presence of the coloured filters adds even more possibilities: it should be possible to have the two telescopes observe the same tile simultaneously but using different filters, to get immediate colour information on all sources across the field. It might also be desirable to accept the impact on survey cadence and have each telescope carry out an independent survey in different filters, or perhaps have one taking rapid 60~s exposures while the other surveys the sky more slowly but to a greater depth. These decisions will ultimately be made depending on the science requirements of the GOTO collaboration, but it is important that the future G\nobreakdash-TeCS scheduling system should be able to handle whatever strategy is chosen.

\subsection{Controlling multiple sites}
\label{sec:multisite}

\begin{figure}[t]
    \begin{center}
        \includegraphics[width=0.95\linewidth]{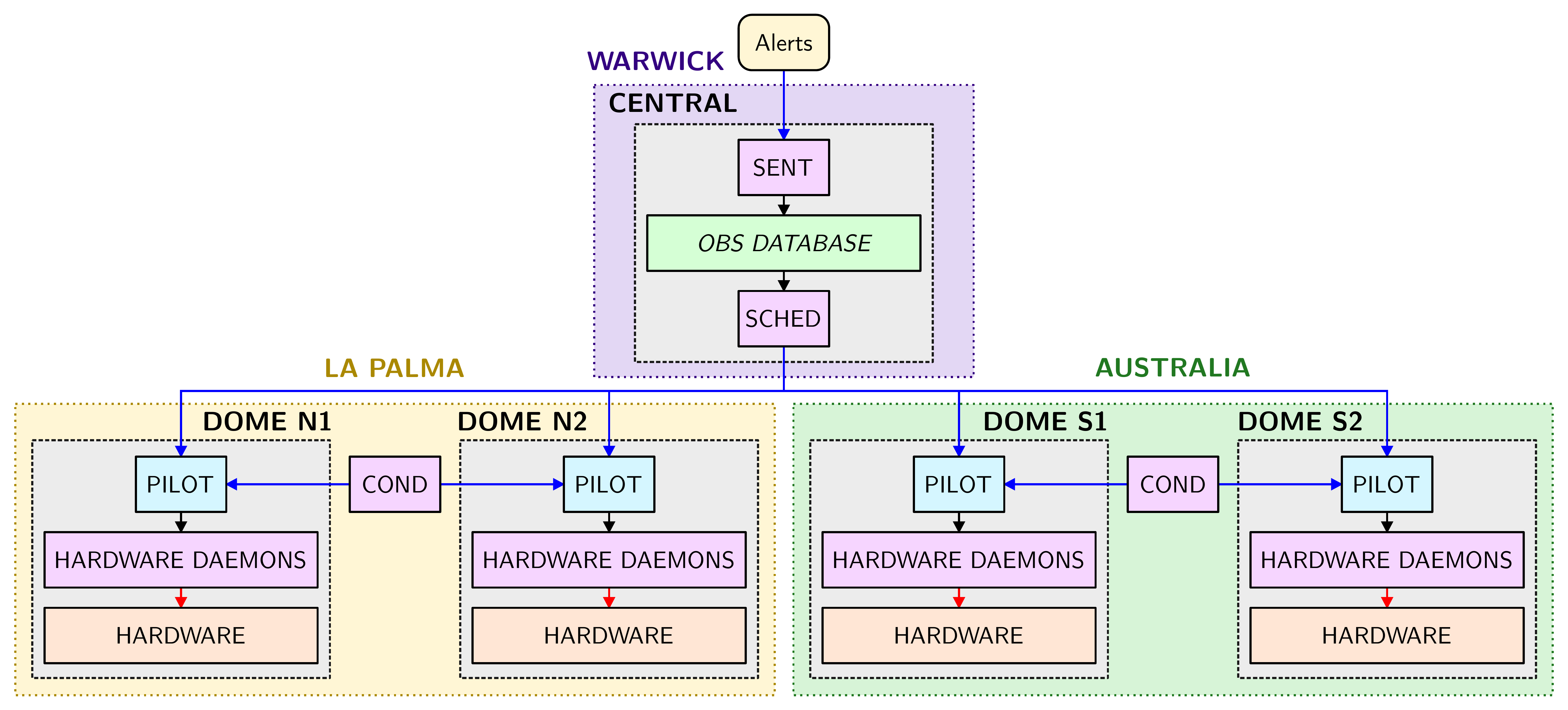}
    \end{center}
    \caption{The final G\nobreakdash-TeCS software architecture, with multiple telescopes at multiple sites each receiving targets from the central scheduler.}
    \label{fig:gtecs_flow_future2}
\end{figure}

As described in Section~\ref{sec:goto}, the GOTO project will ultimately include sites in both the northern and southern hemispheres. The same principle described previously for controlling multiple telescopes will apply wherever they are situated in the world: each will have an independent pilot which will receive targets from a central scheduling system. Figure~\ref{fig:gtecs_flow_future2} expands Figure~\ref{fig:gtecs_flow_future} to cover four telescopes spread across two different sites: La Palma in the Canary Islands and Siding Spring in Australia. The server hosting the observation database and scheduler is currently situated on La Palma, but it is anticipated that this will be relocated to the GOTO data centre at Warwick University, UK, where it will be accessible from both sites.

When upgrading the scheduler to account for multiple sites it will have to take into account not just different horizons but also the position of the Sun, to know which telescopes are available to observe. For the planned GOTO sites there is a fortunate simplification: as the Canary Islands and eastern Australia are near-antipodal, there will never be a case when both sites are observing at the same time (when the Sun is below the horizon for one it will be above the horizon for the other). This drastically simplifies the calculations for the central scheduler: as it will only ever have to determine targets for telescopes based at the same site, it should not need to include any extra logic to that already required for the two La Palma telescopes described in Section~\ref{sec:multitel}, aside from tailoring its decision based on which site is active at the current time.

Although the two sites will never be observing simultaneously, there is still a benefit to them sharing a scheduler, and in particular having a shared observation database. La Palma is located at $28^{\circ}$ north and Siding Spring is located at $31^{\circ}$ south, meaning each telescope can see approximately 75\% of the sky and 50\% of the sky is visible from both sites. With a shared database, the sites will be able to effectively share the load of observing the all-sky survey tiles within this region. Likewise, if a gravitational-wave probability skymap is visible from both hemispheres whichever site starts observing second (i.e. the site where it was daytime at the time the event was received) should take into account which tiles have already been observed by the other site, and then prioritise those which have not yet been observed. Therefore the best way to utilise both sites efficiently is with a fully-connected system, making the entire GOTO network a single global observatory.

\section{Conclusion}
\label{sec:conclusion}

The GOTO Telescope Control System (G\nobreakdash-TeCS) has been successfully commissioned with the GOTO-4 prototype telescope on La Palma. The software controls all aspects of the robotic telescope, including hardware control, conditions monitoring, alert listening and observation scheduling. In particular, the system allowed GOTO to successfully participate in the third LIGO-Virgo gravitational-wave observing run in 2019-20\cite{Gompertz2020,S190814bv}.

As more telescopes are added to the GOTO network, G\nobreakdash-TeCS will evolve from controlling a single telescope to overseeing the entire global network. Each GOTO node will run semi-independently, with a central scheduler prioritising observations and assigning targets to each telescope. By the start of the next gravitational-wave observing run GOTO should be a fully operational network, providing near 24-hour coverage of the visible sky and able to begin follow-up observations of any new detections in less than a minute.

\section*{Acknowledgements}
 
The Gravitational-wave Optical Transient Observer (GOTO) project acknowledges the support of the Monash-Warwick Alliance; the University of Warwick; Monash University; the University of Sheffield; the University of Leicester; Armagh Observatory \& Planetarium; the National Astronomical Research Institute of Thailand (NARIT); the Instituto de Astrofísica de Canarias; the University of Turku; the University of Manchester; the University of Portsmouth and the UK Science and Technology Facilities Council (grant numbers ST/T007184/1 and ST/T003103/1).

\noindent This project makes use of Astropy, a community-developed core Python package for Astronomy\cite{astropy13,astropy18}.

\bibliography{report}
\bibliographystyle{spiebib_shortlist}


\end{document}